\documentclass{amsart}

\usepackage{graphicx}
\usepackage{amsmath}
\usepackage{amssymb}
\usepackage{amsthm}
\usepackage[margin=1in, top=1.5in]{geometry}
\usepackage{booktabs}
\usepackage{float}

\usepackage[slantedGreek]{mathptmx}
\usepackage{url}
\usepackage{bbm}
\usepackage{accents}

\newlength{\dhatheight}

\begin{document}


\title{Self-sustained price bubbles driven by Bitcoin innovations and adaptive behavior}

\author{Misha Perepelitsa and Ilya Timofeyev}

\date{\today}
\address{
Department of Mathematics\\
University of Houston\\
4800 Calhoun Rd.\\
Houston, TX. email: mperepel@central.uh.edu, ilya@math.uh.edu}

\begin{abstract}
We show that infinite divisibility of a trading commodity leads to a self-sustained price bubble when traders use adaptive investment strategies. The adaptive strategy can be viewed as a psychological response of a trader to the situation when the trader's estimation of future prices does not match the actual, realized price.  

We use a multi-agent model to illustrate the price bubble formation
and to quantify its main statistical properties such as the return, the volatility, and the systematic risk of the price bubble to crash. We discuss the plausibility for  bubbles to drive  prices of digital currencies. 
\end{abstract}

\maketitle






\section{Introduction}

With an introduction of Bitcoin in 2010, digital currencies became a global social phenomenon and they are gaining popularity with investors. One of the distinctive features of digital currencies is their ultimate divisibility, as virtually any fraction amount of it can be traded. Currently, the minimum trading amount for Bitcoin is set at $10^{-6}$ units, but the design of the currency allows it to be lowered by any order of magnitude. This, technically, solves the problem of liquidity, as no matter how high the price is, one can always buy some of it, say, with 1 dollar. 
And it seems that the divisibility indeed comes handy, as the price for Bitcoin skyrocketed in the last decade. It's growth is attributed to many factors such as the novelty effect, the demand from the black market, global access to the currency through the internet, the usage of bitcoin as storage in markets with hyper-inflated national currencies,  and, not least, the fact that bitcoin became a sort of a social cult of a decentralized currency, see Dodd (2017).
Apart from the black market demand, which alone can not contribute to this phenomenal growth, all other factors point at a price bubble, which seems to be a consensus among the economists, see  an article by Wolff-Mann (2018).

In this paper we show that this unique property of a traded commodity to be {\it infinitely divisible} makes possible the emergence of {\it self-sustained} price bubbles. By this we mean  that a group of investors can ``pump'' the price up at a positive rate, without ever running out of cash. More than that, we show it may happen without  any collusion between the investors, if each of them follows a psychologically appealing, adaptive investment strategy. 

It is hard to estimate if a speculative bubble of this kind is behind the growth of the Bitcoin.
Several mechanisms a likely to be at work here. For example, the bubble might have started as an adaptive speculative bubble that later morphed into a spontaneously organized, global Ponzi scheme, or the other way around.
Other mechanisms of price bubbles in digital currencies are possible, see for example Griffin and Shams (2020), 
or, for generic models of bubbles, Sornette (2003).  It is certainly not the goal of this work to fit the parameters of the model to the time series of Bitcoin prices and to estimate its statistical significance. Rather than that, our intention was to show the potentiality of  this sort of bubble to develop in the market of digital currencies.

Let us mention that the adaptive behavior of investors is not limited to trading of digital currencies but rather is an ubiquitous phenomenon. Thus the model for the price bubble dynamics, considered in this paper, has a wider applicability. In particular, it shows that for ``classical commodities'' there must be a continuous in-flow of investments for a price bubble to persist, and provides a analytical tool to obtain quantitative information on the operation of a price bubble. 
 

The paper organized as follows. In section \ref{ASPP} we describe the model and its main properties and provide numerical estimates of the return and the volatility of the price.

In section \ref{Risk}, we give numerical estimates of the systematic risk for the price bubble to crash, measured by two different metrics, one based on the low cash-to-stock portfolios of traders, and the other, on low current-cash-to-initial-cash ratio.   We obtain two different estimates of the expected stock price adjusted by the risk of a crash. 

We argue that the second metric is a more relevant measure of risk for the adaptive speculative bubbles. This gives another evidence of a bubble being self-sustained. On the other hand, we argue that  overestimation of the risk by the first metric and the high volatility of the returns may contribute to longevity  of speculative bubbles generated by this mechanism in efficient markets. 

Additionally,  we describe a relation between the volatility of the price and the  systematic risk, see formula \eqref{eq:gamma} on page \pageref{eq:gamma}. It allows for the direct estimation of the systematic risk from an empirical values of the volatility of the price time series.  Several interesting statistical laws of a price bubble in a fixed ``psycho-state'' are derived from this formula, such as the distribution of the times of a crash, and the maximum expected price that can be attained in a life-time of a bubble.

In the section \ref{Conclusions} we give a psychological interpretation of two main parameters of our model and consider a possibility that these parameters have cyclic dynamics in the domain of the psychological variables, thus defining an intrinsic life span of a speculative bubble. This view of a price dynamics as an output of a cyclic model suggests that the price dynamics generated by this model can be in agreement with the no-arbitrage principle.

Models of adaptive economic behavior were developed by many authors in many different contexts. In  strategic decision making  they were advanced as  models of convergence by Cross (1983), models of learning in games by Fudenberg and Levine (1998), and reinforcement learning models by Roth and Erev (1995), to mention just a few contributions.   Our work is close to  the methodology of multi-agent dynamic simulations  of Stigler (1964),  Kim and Markowitz (1989), Arthur et al. (1997), Levy et al.  (1994, 1995, 2000), Lux (1995, 1998), Lux and Marchesi (1999, 2000).  Price bubbles in efficient markets is a well-known phenomenon. They were discussed, for example, by Mandelbrot (1966), Blanchard and Watson (1982) and Sornette (2003).

\section{Asynchronous stochastic price pump}
\label{ASPP}
Adaptive behavior is an universal paradigm. Not surprisingly it is widely used as an investment strategy. Consider for a example an investor who has investment in a stock and has certain cash reserve and would like to keep the stock-to-cash ratio at certain level, so that the investor is  regularly buying or selling stocks to be at that target level. We let our investor, perhaps a bitcoin enthusiast, to be somewhat uncertain about the investment goals, and in particular, to have infinite investment horizon. This might be rather natural assumption in the case of the digital currencies, since they have been introduced just recently. The investor  might consider the market as a complex, exogenous environment with constantly changing characteristics. As an adaptive strategy the investor may increase his investment in stock if the market outperformed the investor's expectations as measured by the stock-to-cash ratio and decrease the investment, otherwise. For example, a trader may increase the target stock-to-cash ratio by a factor of $\alpha>1,$ in the former case, and decrease it by the factor of $\beta<1,$ in the latter. Parameters $\alpha$ and $\beta$ express a degree of the psychological response of the investor to the market performance and can be thought of as measures of investor' optimism and pessimism, or, greed and fear of losing.

What will happen if all other investors in the market follow similar strategies, trying to adapt to the changing market environment that they themselves create? One might expect some sort of anomaly to arise. The scenario was considered by Perepelitsa and Timofeyev (2019) based on the analysis of a multi-agent model, called asynchronous stochastic price pump (ASPP) that we describe below.

\subsection{2-agent model} To introduce the model we start with a market of only two agents trading shares of a single asset. The agents, do not know the asset fundamental value (and maybe there isn't), but expect that the price can grow at moderate rates for long periods of time, and they always prefer increasing values to decreasing.

The agents meet at time intervals that we label with $n=0,1,2..$ At period $n,$ the first agent's portfolio has   $(s^n_1,b^n_1),$  dollar value of stock and bond investments,
and the second: $(s^n_2,b^n_2).$ The stock price per share at period $n$ is $P^n.$ The next period investment portfolio $(s^{n+1}_i,b^{n+1}_i)$ and price $P^{n+1}$ determined trough the following steps.

Evaluating the last change in their portfolios, agents set target stock-to-bond ratios for the next period, that we denote by $k^{n+1}_i.$ The new price $P^{n+1}$ is set in such a way that the dollar amount of funds that agent 1 wishes to move from stocks to bonds, to be at the target ratio $k^{n+1}_1,$ equals the dollar amount agent 2 wants to move from bonds to stock to be at ratio $k^{n+1}_2.$ This balance is expressed as
\begin{equation}
\label{balance}
\frac{P^{n+1}}{P^n}s^n_1{}-{}k^{n+1}_1b^n_1{}={}k^{n+1}_2b^n_2{}-{}\frac{P^{n+1}}{P^n}s^n_2.
\end{equation}
Here we assume that any fractional amount of a share can be exchanged, and for simplicity of the exposition, consider the case with zero interest in the bond account. With the newly set price the agents re-balance portfolios
and evaluate the market performance by comparing ratios $P^{n+1}s^{n}_i/(P^nb^n_i)$ to target values $k^{n+1}_i.$ If $P^{n+1}s^{n}_i/(P^nb^n_i)>k^{n+1}_i,$ from the agent $i$ point of view, the market performed better than expected, and if $P^{n+1}s^{n}_i/(P^nb^n_i)<k^{n+1}_i$ it performed worse.  Since while one agent is selling shares the other one is buying,  they have different opinions on the market performance. The agents update the target portfolio ratios according to the rule
\begin{equation}
\label{eq:feedback}
k^{n+2}_i{}={}\left\{
\begin{array}{ll}
\alpha k^{n+1}_i & \dfrac{P^{n+1}s^{n}_i}{P^nb^n_i}>k^{n+1}_i,\\
\\
k^{n+1}_i & \dfrac{P^{n+1}s^{n}_i}{P^nb^n_i}=k^{n+1}_i,\\
\\
\beta k^{n+1}_i &\dfrac{P^{n+1}s^{n}_i}{P^nb^n_i}<k^{n+1}_i.
\end{array}
\right.
\end{equation}
If the agent identifies a growing market she increases her stock-to-bond ratio by a fixed amount, while the other agent decreases her ratio. The feedback reflects the fact that when faced with a series of bad investments the agent will reduce the equity part of the portfolio, while with the investment growing better than expected, the agent will take riskier position. 

\begin{figure}
\centering
\includegraphics[scale=0.65]{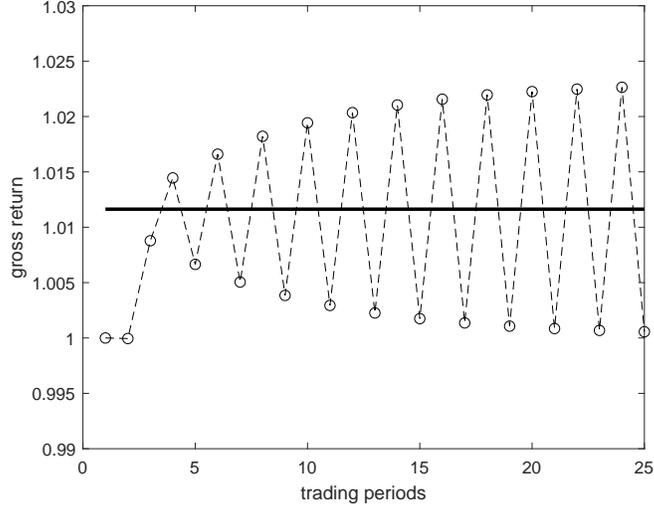}
\caption{Gross return for first 25 trading periods in a 2-agent ASPP model, with $\alpha = 3.01,$ and $\beta=0.36.$ Circles indicate the values of gross returns for alternating ups and downs of stock price, with a net return per period
approaching value $\sqrt{\alpha\beta}=1.012,$ (solid line).}
\label{fig:return_2_agents}
\end{figure}

It easy to see that there is an equilibrium when no shares are traded and price doesn't change: $s^n_i=s^0_i,$ $b^n_i=b^0_i,$ $k^n_i=s^0_i/b^0_i,$ $P^n=P^0.$ Small deviations will set off a non-trivial dynamics. In this process, agents will alternatively change their preferences by factors $\alpha$ and $\beta,$ so that, on average, the stock-to-bond  ratios change at the rate $(\alpha\beta)^{1/2}$ per period. The gross return, $P^{n+1}/P^n,$ after a short transient period will settle at the average value of $(\alpha\beta)^{1/2}$ per period, but will keep oscillating around that value,
as in  Figure \ref{fig:return_2_agents}. The stock price goes up when agents have bias toward the increase of the risk preferences, expressing the optimistic outlook, quantified by the condition $\alpha \beta>1.$ 

The story of this process is simple: agents are trading stock and
and move their stock-to-bond ratios up and down with an upward average trend. This  drives up the average demand and consequently the price.


\subsection{Properties of $N$-agent ASPP}
\label{Formulas}
For a mulit-agent model, we populate the market with many copies of agents considered above  and let them trade anonymously, by submitting their buy-sell orders to a market maker.  Thus,
 we consider a set of $N$ agents, described by their portfolios $(s_i,b_i),$ $i=1\ldots N,$ of dollar values of a stock and cash accounts, and let $k_i$ stand for agent $i$ stock-to-bond ratio, either actual, or a future, target value. $P_0$ will denote a current price per share and $P$ the new price determined by agents' demand. Let $\{i_l\,:\,l=1\ldots m\}$  be the set of  ``active'' agents, i.e. the ones setting the new price. The set of active traders is determined each trading period by a random draw from the population.

If $x_{i_l}$ is the dollar amount that agent $i_l$ wants to invest in stock, then
\[
\frac{\frac{P}{P_0}s_{i_l}+x_{i_l}}{b_{i_l}-x_{i_l}}{}={}k_{i_l}.
\] 
 The demand-supply balance is
\[
\sum_{l=1}^m x_{i_l}{}={}0,
\]
which can be solved for $P:$
\[
\frac{P}{P_0}{}={}\left( \sum_{l=1}^m\frac{k_{i_l}b_{i_l}}{1+k_{i_l}}\right)\left(\sum_{l=1}^m\frac{s_{i_l}}{1+k_{i_l}}\right)^{-1}.
\]
Once the price is set, agents move corresponding amounts between cash and stock accounts, re-balancing their portfolios. The update mechanism for new stock-to-bond ratios is given by \eqref{eq:feedback}.  The interaction is repeated the following trading periods with new, randomly selected sets of active agents. In this trading scheme the total amount of cash in the system and number of shares of stock remain constant.

\begin{figure}
\centering
\includegraphics[scale=0.7]{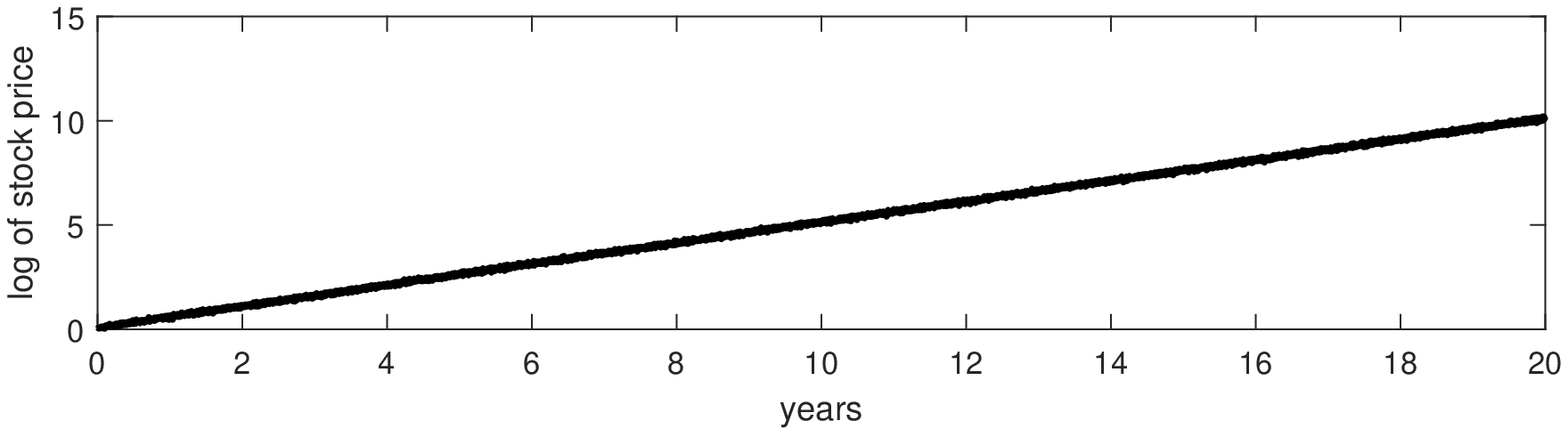}
\includegraphics[scale=0.7]{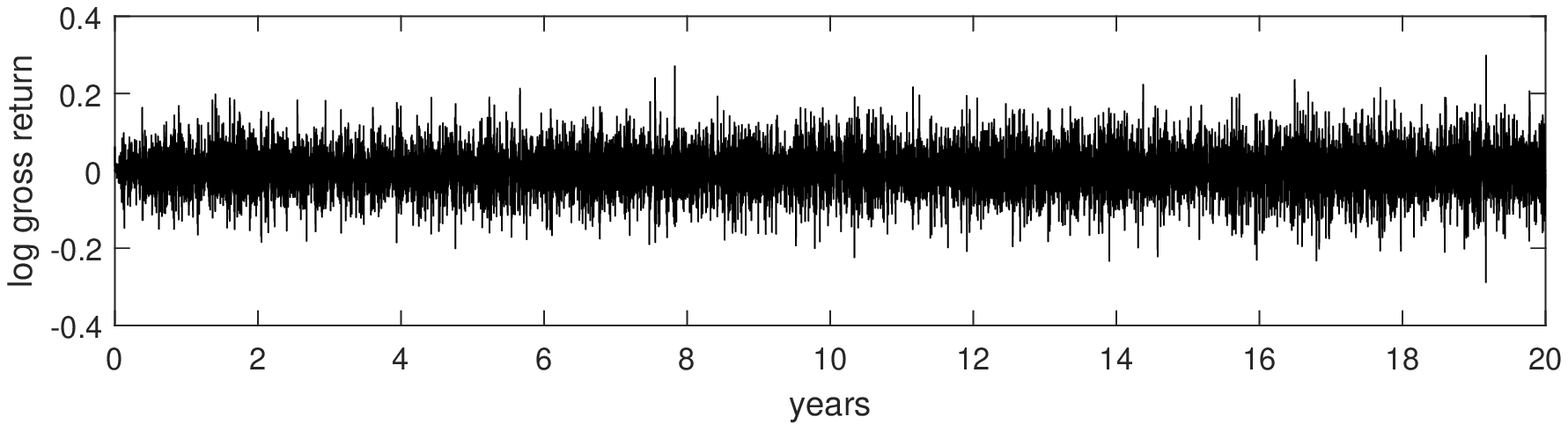}
\includegraphics[scale=0.7]{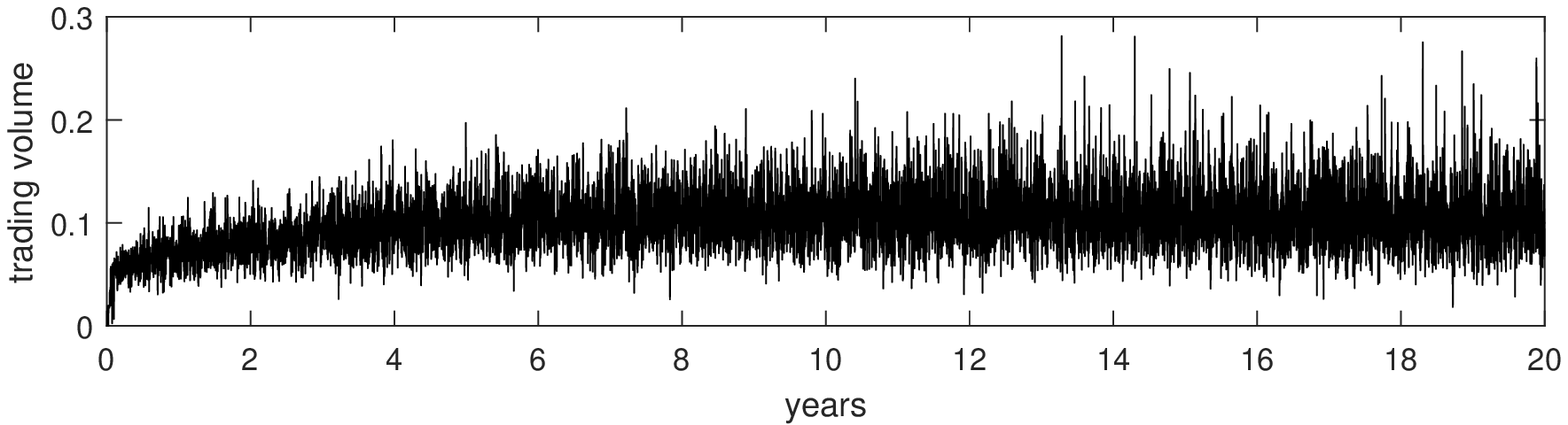}
\includegraphics[scale=0.7]{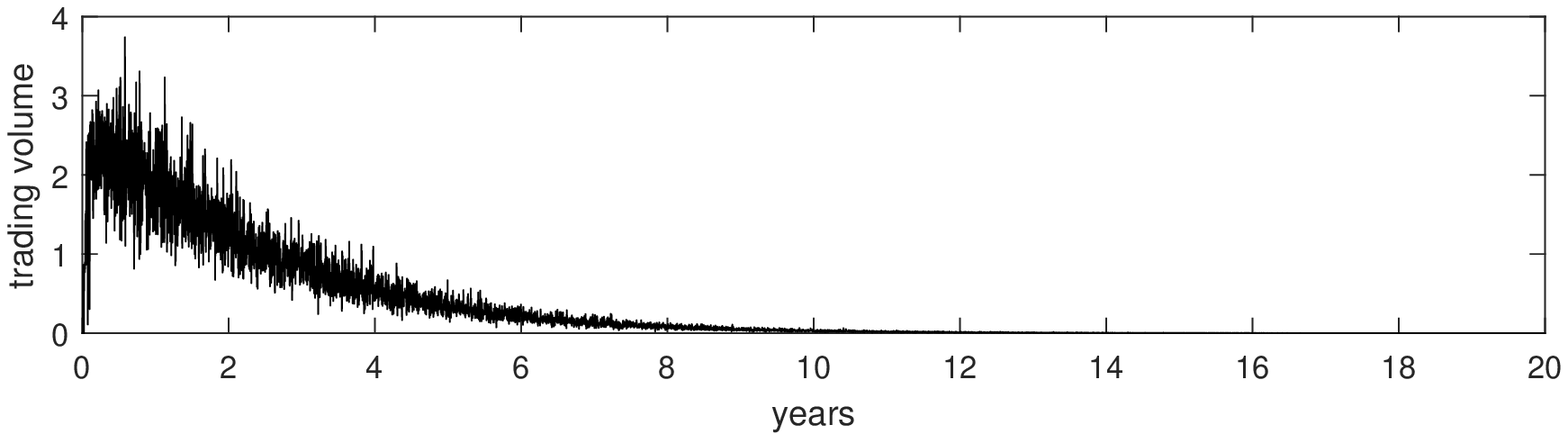}
\caption{Four typical time series describing one simulation of ASPP for a period of 20 years. From top to bottom: logarithm of the stock price, logarithm of the daily gross returns, daily trading volume as percentage of the total cash in the system, daily trading volume in units of shares.}
\label{fig:trajectory}
\end{figure}

There is an equilibrium state, when all agents have balanced portfolios, $s_i/b_i=k_i,$ $i=1,\ldots,N.$ In this case stock price doesn't grow, $P_{n}=P_{n-1}.$ When the parameters  are out of the equilibrium, even by a small degree,  the system exhibits non-trivial, divergent dynamics, such as shown in Figure \ref{fig:trajectory}.

In this divergent regime, after a certain transient period, the logarithm of the price start fluctuating as normally distributed white noise with linearly growing in time mean, see Perepelitsa and Timofeyev (2019). The distribution of returns depends on the mechanism for selecting a random group of active agents. If the number of active agents, $m,$ is fixed the returns in the stationary regime follow log-normal distribution. If number $m$ is itself random, for example chosen from an uniform distribution, the returns have significantly more mass at the tails and at 1, than a corresponding log-normal distribution.

\begin{figure}
\centering
\includegraphics[scale=0.65]{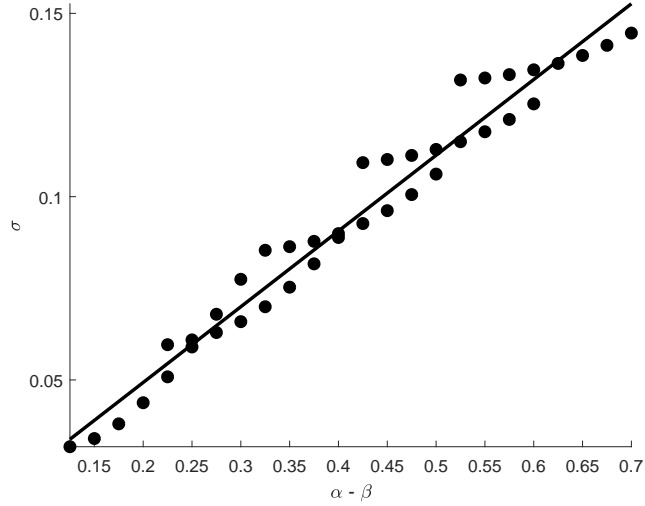}
\caption{ Values of sigma from Table \ref{tab:sigma} plotted against $\alpha-\beta.$  }
\label{fig:sigma}
\end{figure}

To estimate the values of $\sigma$ in \eqref{eq:return2} 
we performed numerical simulation of ASPP model, with $N=500$ and $m=10,20,40,$ on 200,000 trajectories for each pair  of test  values of $(\alpha,\beta).$ The values of $\sigma$ for the case with  $m=10$ active traders are reported in Table \ref{tab:sigma}. Figure \ref{fig:sigma} shows the scatter plot of those values against the values of $\alpha-\beta$ and the corresponding linear regression line.  Linear regression coefficients are listed in Table \ref{tab:reg}. It can be seen from that table that formula $c(m)(\alpha-\beta)$ is a good approximation for volatility $\sigma,$ as a function of $\alpha$ and $\beta.$ In all three cases, there is a positive correlation of about $0.7$ between the mean log-return and the volatility.  

\begin{table}[H]
\centering
\begin{tabular}{@{}lrrrrrrrr@{}}
\hline\noalign{\smallskip}
$\alpha\, \backslash \,\beta$ &   0.8     &  0.825 & 0.85 & 0.875 & 0.9 & 0.925 & 0.95 & 0.975\\
\noalign{\smallskip}\hline\noalign{\smallskip}
1.1      &0.0775&0.0679&0.0590&0.0508&0.0438&0.0380&0.0340&0.0318  \\
1.2      &0.0889&0.0817&0.0753&0.0700&0.0659&0.0629&0.0609&0.0596  \\
1.3		 &0.1062&0.1066&0.0962&0.0927&0.0899&0.0878&0.0864&0.0854  \\
1.4      &0.1253&0.1211&0.1117&0.1150&0.1129&0.1112&0.1101&0.1093 \\
1.5      &0.1446&0.1413&0.1385&0.1363&0.1346&0.1333&0.1324&0.1318\\
\hline\noalign{\smallskip}
\end{tabular}

\vspace{15pt}

\caption{$\sigma$ as a function of  $\alpha$ and $\beta$ for ASPP model with $N=500$ and $m=10,$ in a stationary regime reached after 2 years of trading time.
 \label{tab:sigma}}

\end{table}

\begin{table}[H]
\centering
\begin{tabular}{@{}lcc@{}}
\hline\noalign{\smallskip}
    &  c     &  d \\
\noalign{\smallskip}\hline\noalign{\smallskip}
m=10      &0.2066&0.0081 \\
m=20      &0.1457&0.0055  \\
m=40		 &0.1084&0.0027 \\
\hline\noalign{\smallskip}
\end{tabular}

\vspace{15pt}

\caption{Linear regression coefficients $(c,d)$ in  $\sigma=c(\alpha-\beta) + d$ for several values of $m.$ 
 \label{tab:reg}}
\end{table}

Combining these findings and assuming for the simplicity of presentation that there is one trade per day (total 360 per year), and using $t_n=n/360$ to label $n^{th}$ trading period on the time axis measured in years,  we obtain that daily prices $P(t_n)$ are independently distributed with  
\begin{equation}
\ln(P(t_n))\in \mathcal{N}(n\ln r,\,\sigma/\sqrt{2}), \quad  
\label{eq:return}
r{}={}(\alpha\beta)^{\frac{m}{2N}}, \quad \sigma=c(m)(\alpha-\beta).
\end{equation}
In this formulas, $N$ is the total number of investors and $m$ is the number of investors participating in one trade per day, and $c(m)$ is a positive number, depending on $m.$  Accordingly, for the daily returns, $R(t_n)=P(t_n)/P(t_{n-1}),$ we have 
\begin{equation}
\label{eq:return2}
\ln(R(t_n))\in  \mathcal{N}(\ln r,\,\sigma)
\end{equation}
where $r$ and $\sigma$ as in \eqref{eq:return}.

Let's illustrate this behavior on a particular model setup, in which there is total of $N=500$ investors, with $m=10$ randomly chosen agents to participate in a daily trade, and the value of optimism is set at $\alpha=1.2$ and pessimism at $\beta=0.96$ 
Figure \ref{fig:trajectory} illustrates typical characteristic features of a stochastic price pump. 
The top panel shows time series of the logarithm of the price of a share with a ``healthy'' growth rate $1.66$ per year (66\% increase). The panel below shows that the returns are not constant but fluctuate as a white noise with volatility of about $0.07$ (7\% daily price changes). This price growth is achieved by a  modest amount of trading, with daily trading volume not exceeding $0.3\%$ of the total cash in the system, see the third panel from the top in Figure \ref{fig:trajectory}. Combined with the exponential growth of the price, this means that the daily trading volume measured in the units of the shares progressively decreases, see the bottom panel in Figure \ref{fig:trajectory}. If, initially, trades evolved exchanges of amount of shares of order 1, then by year 10, the typical daily volume is of order $10^{-2}$ and by year 20, of order $10^{-4}.$ This property is a consequence of the exponential growth of the price combined with the fact that the trading conserves the amount of cash in the system.

This example shows that investors acting adaptively  can generate a long-lasting, positive price dynamics, and a necessary condition for an ASPP to function for prolonged time is an infinite divisibility of traded commodity, which is one of the main characteristic properties of digital currencies. 

Our main interest in this paper will be the systematic risk of ASPP and its relation to the volatility, which we discuss in the next section.

\section{Systematic risk}
\label{Risk}
Multi-agent models offer a convenient tool for estimating systematic risk of sharp price drops (crashes), since the information about investment portfolio of each agent is a part of the description of the model. Here, by systematic risk we understand the risk of a breakdown of the system that generates the price dynamics, in this case, the breakdown of ASPP. We consider two metrics for measuring this risk. The first is based on the proportion of agents with low cash-to-stock ratios. Let us denote by $q(t)$ the proportion of agents with cash-to-stock ratio less than, say 10\%, at time $t.$ Higher values of $q(t)$ mean higher likelihood of a crash, as a larger proportion of agents may spontaneously initiate stock sellout. We will use a hazard rate function
\begin{equation}
\label{hq}
h_q(t){}={}\frac{q(t)}{1-q(t)}
\end{equation}
to quantify the probability of a crash. According to this formula, the hazard rate changes monotonically from $0,$ when $q(t)=0,$ to $+\infty,$ when $q(t)=1.$ Recall that the hazard rate function is defined so that for an infinitesimal time interval of length $dt,$
$h_q(t)dt$ is the probability of a crash to occur during interval $[t,t+dt],$ given that it didn't occur before time $t.$ Then, the probability that no crash occurs up to time $t,$ equals
\begin{equation}
\label{Sq}
S_q(t) = e^{-\int_0^th_q(s)\,ds}.
\end{equation}
Thus, the ``discounted'' stock price at time $t$ equals $P(t)S_q(t),$ assuming, for simplicity, that a crash brings price to zero. 

Alternatively, the systematic risk can be computed based off the proportion of agents with low ratio of cash reserves at time $t$ to the initial cash reserves, say less than 60\%. We denote this proportion by $r(t),$ and the corresponding hazard rate and the survival functions by $h_r(t)$ and $S_r(t),$  defined by \eqref{hq} and \eqref{Sq} where $h_q$ is replaced by $h_r.$

\begin{figure}
\centering
\includegraphics[scale=0.75]{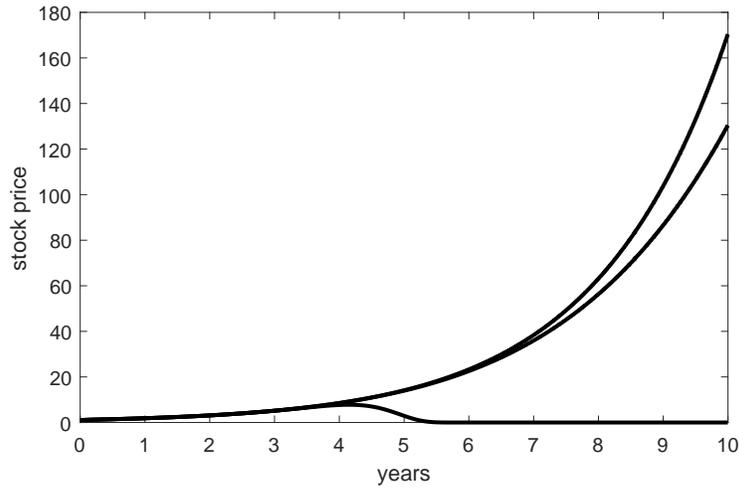}
\caption{Expected stock price discounted by the risk of a crash. Top graph is  the expected price without discounting. The graph in the middle is the expected price discounted by hazard rate $h_r(t).$ The graph at the bottom is the expected price discounted by hazard rate $h_q(t).$ }
\label{fig:discounted}
\end{figure}

Figure \ref{fig:discounted} shows plots of the expected price, and the prices discounted by two hazard rates, $h_q$ and $h_r,$ for the same model as used in the simulation  in Figure \ref{fig:trajectory}. Measured by the  first metric, the price collapses to zero, while in the second metric the price continues to grow (eventually it will go down as well).

It might seem that  the first metric seems to be more relevant measure of a sudden sellout. Note however, that  a low cash-to-stock ratio is not caused by traders spending all cash buying stock, but rather on the anomalous price growth. For this reason the second metric is more likely to be used by adaptive speculators with long investment horizon.  As explained in the previous section the latter is model of main interest in this paper.
 
Another difference between two metrics, that we would like to emphasis here, is that the first depends on the initial investment portfolios, and can be arbitrary high at time $t=0$ because of that. Whereas  the second is always zero at $t=0$ and increases gradually in the process of trading, thus being a characteristic of a fully developed stationary regime of the price pump.

We can conclude that the persistence of the price bubbles in the digital currencies may result from overestimation of the risk by either the hazard rate $h_q(t),$ whereas  the actual risk in an adaptive speculation model given by the hazard rate $h_r(t)$ with significantly smaller values, or the high volatility which is positively correlated with the return, see formulas for $r$ and $\sigma$ in \eqref{eq:return}. The latter might be the reason why short term investors stay away from them.

As can be seen from the previous examples,  for fixed values of $\alpha$ and $\beta,$ the volatility of the price is constant. At the same time, the hazard function  increases, implying that the volatility can not be a correct measure of the systematic risk. However, they are not completely unrelated. To show this, let us define a cumulative volatility at $t$ years as 
\[
V(t) = \int_0^t \sigma^2(s)\,ds.
\]
For a time series of daily stock price $P(t_n),$ the cumulative volatility  $V(t_n) = \sum_{t_m\leq t_n} [\ln R(t_m) - \ln R(t_{m-1})]^2/720,$ where $R(t_n)=P(t_n)/P(t_{n-1})$ is the gross return. 

\begin{figure}
\centering
\includegraphics[scale=0.7]{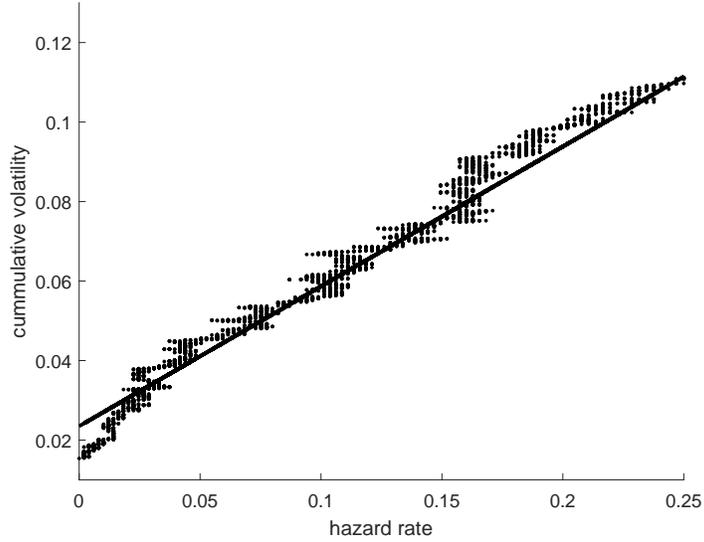}
\caption{Scatter plot for cumulative volatility $V(t_n)$ and hazard rate $h_r(t_n),$ in one simulation of the model for the first 20 years.}
\label{fig:vol_hazard}
\end{figure}

Figure \ref{fig:vol_hazard} shows a scatter plot of daily values of $V(t_n)$ and the hazard rate $h_r(t_n),$ and the best fit linear approximation (linear regression). This numerical evidence is strongly suggestive of a linear relation between $V(t)$ and $h_r(t),$ or, equivalently, a linear relation 
\begin{equation}
\label{eq:gamma}
\sigma^2 = \gamma \frac{dh_r}{dt}.
\end{equation}
Table \ref{tab:gamma} lists the statistical values of parameter $\gamma$ for several values of $(\alpha,\beta).$ This formula provides an important practical tool for estimation of the systematic risk from the volatility, since the latter can be computed from time series of the price.   
 
From this formula we can find the hazard rate $h_t(t) = \sigma^2 t\gamma,$ and  the PDF for the time of a crash. The latter has a Weibull distribution
\[
\frac{\sigma^2t}{\gamma}{\rm exp}\left({-\frac{\sigma^2}{2\gamma}t^2}\right).
\]
Whith that, the expected discounted stock price becomes equal
\[
P_d(t) = P(0){\rm exp}\left( r_0 t -\frac{\sigma^2}{2\gamma}t^2\right), \quad r_0=\frac{360m}{2N}\ln(\alpha\beta).
\]
It can be seen that there is a maximum value  
$\max  P_d(t)$ of the price, attained when $t= 2r_0\gamma/\sigma^2.$ It places a limit on the price  that  a bubble can reach in a fixed ``psycho-state'' $(\alpha,\beta).$

\begin{figure}
\centering
\includegraphics[scale=0.75]{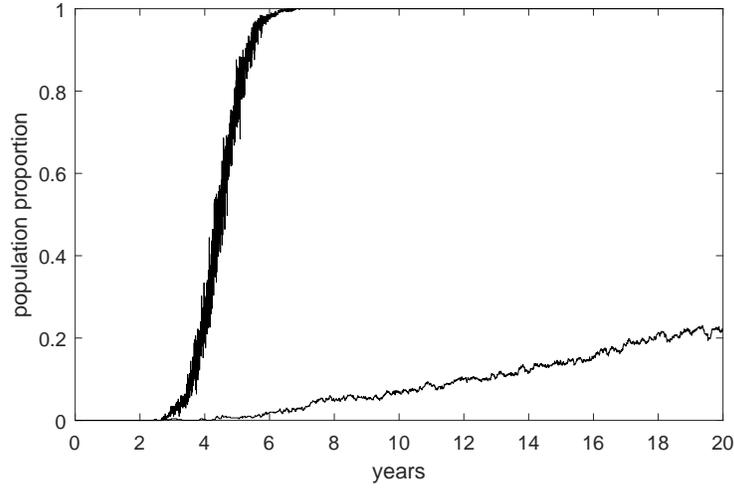}
\caption{Proportions of traders with low cash-to-stock portfolios (left) and low cash-to-initial-cash ratio (right) as functions of time in one simulation of the model for the period of 20 years.}
\label{fig:proportions}
\end{figure}

\begin{table}[H]
\centering
\begin{tabular}{@{}lrrr@{}}
\hline\noalign{\smallskip}
$\alpha\,\backslash\,\beta$ & 1	       & 0.99     &  0.96\\
\noalign{\smallskip}\hline\noalign{\smallskip}
1.1      &  0.2909  &  0.3932  &   0.8343  \\
1.2     &  0.2591  & 0.2763   &   0.3577  \\
1.3		 &  0.3644  & 0.3732   &   0.4084  \\
\hline\noalign{\smallskip}
\end{tabular}

\vspace{15pt}

\caption{Values of $\gamma$ in formula \eqref{eq:gamma}.  Each value is an average of the linear regression coefficients of $V(t_n)$ on $h_r(t_n),$ for 1000 20-year path simulations of ASPP model with  $N=500$ agents in the system, $m=10$ active agents.
 \label{tab:gamma}}

\end{table}

\section{Conclusions}
\label{Conclusions}
In the last section we listed several factors that can contribute to  a prolonged life span of the price bubble. But will the bubble perpetuate its growth forever? What will happen in an ideal  case when no crashes due to systematic risk occur, when there are no arbitrageurs out there, and investors have very long, or even infinite, investment horizon?  The answer to these questions lies in the interpretation of parameters $(\alpha,\beta)$ of the model, that we identify as two psychological motivators, greed and fear, receptively. Here, we are assuming that a coherent psychological state in the population is achieved through the copying behavior and inter-connectedness of the investors by information networks, so that all investors share the same or almost the same values of $\alpha$ and $\beta.$ These parameters are combined together in  the formula for the return \eqref{eq:return}, which shows that the price pump can work both ways, as a price inflator or a deflator.

Being of psychological nature $(\alpha,\beta)$ will change according  to some intrinsic mechanisms, that combine personal psychological response of each investor
to the changes in the composition of his/her portfolio and the degree of influence of this response on the behavior of other investors.  As an approximation, one can assume that $(\alpha,\beta)$ co-evolve in such a way that $\alpha$ (greed) is self-reinforcing and generates $\beta$ (fear), and fear, in its turn, inhibits greed, by  increasing investor's sensitivity to idiosyncratic noise.  

According to this model, the evolution of a price bubble will proceed in a ``predator-prey'' type cycle in the domain of psychological variables, triggered by an exogenous shock. Along the cycle, the return and volatility of the price will changes according to formulas  from \eqref{eq:return}, producing a pattern of  increasing returns with low volatility, followed by 
a period of growing volatility, during which the return reaches its maximum and starts to decrease, which eventually ends with sharp decrease on return and low volatility. At the end of the cycle the system will reach some ``background level'' of $(\alpha,\beta),$ at which they balance each other. The system will remain in this ``no bubble'' phase until another external factor pushes it on a road to a new bubble.

This type of dynamics can produce complex patterns of price variations. Since the period of a bubble cycle is a non-constant  (random) variable, there can be little predictability of the future price, even more so with crashes due to systematic risk. This suggests an intriguing idea that the price fluctuations due to adaptive speculative market behavior might be consistent with the no-arbitrage principle.


%

\end{document}